\documentclass[aps,twocolumn,showpacs,amsmath,amssymb]{revtex4}
\usepackage{graphicx}
\usepackage{dcolumn}
\usepackage{bm}
\begin{document}
\title{Note on the super inflation in loop quantum cosmology}
\author{Kui Xiao}
\email{87xiaokui@mail.bnu.edu.cn}
\affiliation{Department of Mathematical and Physical Teaching, Hunan Institute of Technology, Hengyang 421002, China}
\author{Xiao-Kai He}
\email{hexiaokai77@163.com}
\affiliation{Department of Educational Science, Hunan First Normal University, Changsha 410205, China}
\author{Jian-Yang Zhu}
\email{zhujy@bnu.edu.cn}
\affiliation{Department of Physics, Beijing Normal University, Beijing 100875, China}
\date{\today}
\begin{abstract}
Phenomenological effect of the super-inflation in loop quantum cosmology (LQC) is discussed. We investigate the case that the
Universe is filled with the interacting field between massive scalar field and radiation. Considering the damping coefficient $\Gamma$ as a constant, the changes of the scale factor during super-inflation with four different initial conditions are discussed, and we find that the changes of the scale factor depends on the initial values of energy density of the scalar field and radiation at the bounce point. But no matter which initial condition is chosen, the radiation always dominated at the late time. Moreover, we investigate whether the super-inflation can provide enough e-folding number. For the super-inflation starts from the quantum bounce point, the initial value of Hubble parameter $H(t_i)\sim0$, then it is possible to solve the flatness problem and horizon problem. As an example, following the method of \cite{Amoros-prd} to calculate particle horizon on the condition that the radiation dominated at bounce point, and we find that the Universe has had enough time to be homogeneous and isotopic.
\end{abstract}
\maketitle
\section{Introduction}
The inflation theory is introduced to solve a number of cosmological conundrums (the monopole, horizon, flatness, and entropy problems)
in the standard cosmological model \cite{Lindle}. The inflation stage sits in an area where the Hubble parameter $H$ is approximately a constant, and $\ddot{a}>0$. In this stage the equation of state parameter  is $\omega=-1$,  which is the equation of state for the scalar field usually. It means that the potential of the scalar field is dominated in this stage, until the inflation is ended. At the time $\dot{\phi}$  cannot be ignored any more, and $\omega= -\frac 13$. In the inflation area, the scalar field has no interaction with other field except for the gravitational field, and the density fluctuation is caused by the fluctuation of the vacuum. The inflation stage is lying in a low  temperature area, so this inflation theory is always called cool inflation (or, standard inflation), which is in contrast with warm inflation. Warm inflation is caused by interacting fields between scalar field and radiation. Just like the cool inflation, the warm inflation stays in a stage where the potential of scalar field dominated. A lot of evidences show that the inflation is a brilliant candidate theory to explain the very early  universe (i.e., see \cite{inflation-planck,Lisa-warm}). But there are still some problems need to be solved, one of which is the singularity of the Universe. If the Universe is filled with  radiation and matter, according to the general relativity theory, coming back in time, one concludes that there exists a primeval singularity which is named as the big bang singularity. The big bang singularity problem could be seen as a deficiency of Einstein cosmology at high energies.

To solve the big bang singularity problem, one possible solution is modifying the theory of general relativity at high energies. There are lots of candidates, one of them is LQC (more recently review, please see \cite{Ivan-LQC,Bojowald-book}). LQC is a canonical quantization of homogeneous spacetimes based upon the techniques used in loop quantum gravity (LQG) \cite{Rovelli-Book,Thiemann-Book}. Due to the homogeneous and isotropic spacetime, the phase space of LQC is simplifier than LQG, e.g., the connection is determined by a single parameter called $c$ and the triad is determined by $p$. Recently, it has been showed that the loop quantum effects can be described very well by effective modified Friedmann dynamics.  People always consider two kinds of correction of effective LQC, one is inverse volume correction,  the other is holonomy correction. Considering these modification, one can obtain many interesting results, e.g., the big bang will be replaced by big bounce \cite{Ashtekar}, the most of singularity will be avoid \cite{Singh,Haro-JCAP}, the inflation will be more likely to occur \cite{Ashtekar-probability,Corichi_measure,Linsefors-prediction}, and so on. But the first modification suffers from gauge dependence which cannot be cured and thus yields unphysical effects. We will discuss the dynamical behavior of scalar field in LQC based on the secondary modification. In this effective LQC, a term, $-\frac{\rho^2}{\rho}$, will be added to the right sides of the standard Friedmann equation. Since this correction comes with a negative sign, the Hubble parameter $H$, and then $\dot{a}$ will vanishes when $\rho=\rho_c$, and the quantum bounce occurs and then oscillates forever. Therefore, using the second type of the modification to the Friedmann equation, the physically appealing features of the first one are retained. {Moreover,} for a Universe with a large-scalar factor, the first type of modification to the Friedmann equation can be neglected and only the second type of modification is important.

The super-inflation, which is caused by the quantum geometry effect in LQC \cite{Bojowald-PRL-in}, has been studied by many papers \cite{Copeland-superinflation,Ranken,Ribassin,Amoros-prd}. Some papers found that the scale factor changes too small to solve the flatness and horizon problem during super-inflation stage \cite{Bojowald-PRL-in}, whereas the Hubble parameter $H$ is visibly growing during this stage, then some papers argued that the horizon problem can be solved at the super-inflation area \cite{Copeland-superinflation,Amoros-prd}. Also, the duration of super-inflation, or the change of scale factor depends on the initial value of the scalar field \cite{Mielczarek}, or the parameter of the potential of the scalar field \cite{Ranken}. All those researches discussed the Universe contains a self-interact or massless scalar field. And the simple fluid models for super-inflation also discussed \cite{Ribassin}, and they found that it is possible to distinguish the spectrum of B-modes of super-inflation in LQC and standard inflation in the future observations. As we mentioned at the first paragraph of this section, depending on whether the scalar field can interact with the radiation or not, the inflation theory can be called as warm inflation or cool (standard) inflation theory. The warm inflation theory in LQC has been discussed by \cite{Herrera,Xiao,Zhang}. According to the works of \cite{Herrera,Xiao,Zhang}, the warm inflation is different from the one in Einstein cosmology, so, it is nature to investigate the appearance of super-inflation when considering the scalar field interacts with radiation during this stage. In this Letter, we will discuss this phenomenon.

This Letter is organized as follows: in Sec. \ref{s2}, we will introduce some basic equations of LQC and the warm inflation; in Sec. \ref{s3}, we will show the changes of scale factor during the super-inflation stage, in Sec. \ref{s4}, we will discuss the horizon problem of super-inflation, and we will give some conclusions in the last Sec. \ref{s5}.

\section{\label{s2}Basic equations}

We focus on the flat FRW cosmology in this Letter. The effective equations in LQC are derived from the LQC Hamiltonian constraint operator and include the leading order quantum gravity corrections to the classical Friedmann equation. It turns out that the effective equations provide a surprisingly good approximation to the dynamics of sharply peaked states in LQC at all times, including at the bounce point where quantum gravity effects are strongest \cite{Edward}. In this Letter, we focus on the holonomy correction, then the modified Friedmann equation reads
\begin{eqnarray}
H^2=\frac{8\pi G}{3}\rho\left(1-\frac{\rho}{\rho_c}\right),\label{Fri}
\end{eqnarray}
in which  $\rho_c$ is the critical energy density. We consider the Universe is sourced by a scalar field $\phi$ with positive potential
$V $ coupled with radiation. Then, the total energy density is  $\rho=\rho_\phi+\rho_\gamma$, with the energy density of scalar field $\rho_\phi=\frac12\dot{\phi}^2+V $ and the energy density of radiation $\rho_\gamma$. The potential $V$ is usually considered as a function of the field $\phi$ and the temperature $T$, i.e., $V=V(\phi,T)$, but, for simplicity, we just consider it is a function of the field $\phi$, and we assume potential is a chaotic potential $V(\phi)=\frac12m\phi^2$, with the mass of scalar field $m$.  Considering the conversation of energy density, one can get
\begin{eqnarray}
&\ddot{\phi}+(3H+\Gamma)\dot{\phi}=-V_{,\phi},\label{dphi} \\
&\dot{\rho}_\gamma+3\gamma H{\rho_\gamma}=\Gamma\dot{\phi}^2,\label{drg}
\end{eqnarray}
in which $\Gamma$ is a damping coefficient and is responsible for the decay of the scalar field into radiation. Usually, $\Gamma$ is considered as a function of the field $\phi$, or temperature $T$, or both, or, just a constant. In this Letter, we just consider it is a constant. According to the second law of thermodynamics, $\Gamma>0$ should be satisfied. The values of $\Gamma$ and $m$ should be determined by the observational data, and some connections may exist between them, just as shown in \cite{Herrera}. But we just consider a phenomenological  results of interacting scalar field and radiation during super-inflation stage, so we choose this two parameters arbitrary. $\gamma$ is the adiabatic index, which satisfies $p_\gamma=(\gamma-1)\rho_\gamma$,  for the pressure $p_\gamma$ of radiation,then $\gamma=\frac43$.

Using Eqs.(\ref{dphi}) and (\ref{drg}), it is easy to verify the total energy density is conservation
\begin{eqnarray}
\dot{\rho}+H(3\dot{\phi}^2+4\rho_\gamma)=0.\label{ted}
\end{eqnarray}

Combining Eq.(\ref{Fri}) with Eq.(\ref{ted}), one can get
\begin{eqnarray}
\dot{H}
&=&-\frac{8\pi G}{6}\left(\dot{\phi}^2+\gamma{\rho_\gamma}\right)\left[1-\frac{2}{\rho_c}\left(\frac12\dot{\phi}^2+V(\phi)+{\rho_\gamma}\right)\right].\nonumber\\
\label{dH}
\end{eqnarray}
According to Eq.(\ref{Fri}), we can find that $H=0$ when $\rho=\rho_c$, which is the quantum bounce point. It is easy to find that $\dot{H}>0$ at the bounce point  and it holds positive until $\frac12\dot{\phi}+V(\phi)+\rho_\gamma=\frac12\rho_c$. $\dot{H}>0$ means the universe is in a super-inflation stage. In this stage, the Hubble parameter increases, and the scale factor changes also. For $\frac12\rho_c<\rho<\rho_c$, this stage is staying in quantum effect dominated area, then the super-inflation is rooting in the quantum  geometry effect.

\section{\label{s3}scale factor}

To solve the horizon problem, it needs at least 60 e-folding number. In the standard inflation theory, the scale factor $a$ changes rapidly and the Hubble parameter $H$ remains nearly constant. According to the truth that the e-folding number comes from the changing of the scale factor in standard inflation, someone concluded that the e-folding number created during super-inflation in LQC is not sufficient \cite{Bojowald-PRL-in}, but then it was found that the e-folding number during super-inflation depends on the initial conditions \cite{Mielczarek} or the parameter of the potential of the scalar field \cite{Ranken}. To show the change of the scale factor $a$ during the super-inflation stage, we will discuss the Universe filled with  the interacting field between massless scalar field and radiation at first, and, then give some numerical results in this section.

\subsection{\label{s3A}Massless scalar field}

The massless scalar field is always discussed in LQC. Considering the case $V(\phi)=0$, then the energy density and the pressure of scalar field have the same form ${\rho_\phi}={p_\phi}=\frac12\dot{\phi}^2$, and the equation of evolution of scalar field Eq.(\ref{dphi}) becomes
\begin{eqnarray}
&\ddot{\phi}+(3H+\Gamma)\dot{\phi}=0.
\end{eqnarray}
For simplicity, we consider $\Gamma\gg 3H$ case,then
\begin{eqnarray}
&\dot{\phi}(t)=C e^{-\Gamma t},
\end{eqnarray}
with a integral constant $C$.

It should be mentioned that the super-inflation ends while $\rho=\frac12\rho_c$. Always, scale factor changes very small during the super-inflation stage. It is convenient to assume that $t$ is very small when super-inflation ends, and the duration of super- inflation is very short, then, scale factor can be expanded in Taylor series
\begin{eqnarray}
a=1+a_1t+a_2t^2+a_3t^3+\cdots,
\end{eqnarray}
hence the Hubble parameter becomes
\begin{eqnarray}
H=\frac{\dot
a}{a}=a_1+[2a_2-a_1^2]t+[3a_3-3a_1a_2+a_1^3]t^2+\cdots,\nonumber \\
\end{eqnarray}
with some constants $a_1,a_2,a_3,\cdots$, where we have assumed that  $a(t=0)=1$ at the bounce point. Then the energy density of scalar field can be written as
\begin{eqnarray}
\rho_{\phi}=\rho_{\phi_0}+\rho_{\phi_1}t+\rho_{\phi_
2}t^2+\rho_{\phi_3}t^3+\cdots,
\end{eqnarray}
with
\[
\rho _{\phi _1}=-(6a_1+2\Gamma )\rho _{\phi _0},
\]
\[
\rho _{\phi _2}=(21a_1^2+12a_1\Gamma +2\Gamma ^2-6a_2)\rho _{\phi _0},
\]
\begin{eqnarray*}
\rho _{\phi _3} &=&-\frac 13[(6a_1+2\Gamma )\rho _{\phi
2}+(12a_2-6a_1^2)\rho _{\phi _1} \\
&&+(18a_3-6a_1a_2+6a_1^3-12a_1a_2)\rho _{\phi _0}].
\end{eqnarray*}
in which $\rho_{\phi_0}$ is the energy density of scalar field at the bounce point.

According to Eq. $(\ref{drg})$, it is easy to get the expression of the energy density of radiation
\begin{eqnarray}
\rho_{\gamma}=\rho_{\gamma_0}+\rho_{\gamma_1}t+\rho_{\gamma_
2}t^2+\rho_{\gamma_3}t^3+\cdot\cdot\cdot
\end{eqnarray}
with the initial energy density of radiation $\rho_{\gamma_0}$  at the bounce point and
\[
\rho _{\gamma _1}=2\Gamma \rho _{\phi _0}-3\gamma a_1\rho _{\gamma _0},
\]
\begin{eqnarray*}
\rho _{\gamma _2} &=&-\Gamma \rho _{\phi _0}(6a_1+2\Gamma +3a_1\gamma ) \\
&&+\gamma \rho _{\gamma _0}\left( \frac 92\gamma a_1^2-3a_2+\frac 32%
a_1^2\right) ,
\end{eqnarray*}
\begin{eqnarray*}
\rho _{\gamma _3} &=&\frac 23\Gamma \rho _{\phi _2}-a_1\gamma \rho _{\gamma
_2}-\left( 2a_2-a_1^2\right) \gamma \rho _{\gamma _1} \\
&&-\left( 3a_3-3a_1a_2+a_1^3\right) \gamma \rho _{\gamma _0}.
\end{eqnarray*}

Now we can determine the expanding coefficients $a_0,a_1,a_2,\cdot\cdot\cdot$ of $a(t)$ by the analysis of Eq.$(\ref{Fri})$. More explicitly, we have
\begin{eqnarray}
a_1=\left[\frac{8\pi G}{3}\left(\rho_{\phi_0}+\rho_{\gamma_
0}-\frac{(\rho_{\phi_0}+\rho_{\gamma_0})^2}{\rho_c}\right)\right]^{1/2},
\end{eqnarray}
with
\begin{eqnarray}
&a_1=0\nonumber\\
&a_2=4\pi G(\rho_c-\rho_{\gamma_0}+\frac{\gamma}{2}\rho_{\gamma_0})\nonumber\\
&a_3=\frac{4}{3}\pi G \Gamma
(\gamma-2)(\rho_c-\rho_{\gamma_0})\nonumber
\end{eqnarray}
in which we have used $\rho_{\phi_0}+\rho_{\gamma_0}=\rho_c$ at the bounce point. Then we have the expression of the scale factor $a$ and the
Hubble parameter $H$:
\begin{eqnarray}
a &=&1+4\pi G(\rho _c-\rho _{\gamma _0}+\frac 12\gamma \rho _{\gamma _0})t^2
\nonumber \\
&&+\frac 43\pi G\Gamma (\gamma -2)(\rho _c-\rho _{\gamma _0})t^3+\cdot \cdot
\cdot ,
\end{eqnarray}
\begin{eqnarray}
H &=&4\pi G\left[ 2\left( \rho _c-\rho _{\gamma _0}+\frac 12\gamma \rho
_{\gamma _0}\right) t\right.   \nonumber \\
&&\left. +\left( \gamma -2\right) (\rho _c-\rho _{\gamma _0})\Gamma
t^2+\cdot \cdot \cdot \right] .
\end{eqnarray}

Now, from $\dot {H}=0$, it could be deduced that
\begin{eqnarray}
t_{end}=\frac{\rho_c-\rho_{\gamma_0}+\frac{\gamma}{2}\rho_{\gamma_0}}{\Gamma (\gamma-2)(\rho_c-\rho_{\gamma_0})},
\end{eqnarray}
then
\begin{eqnarray}
a_{end}=1+\frac{16}3{\frac {\pi\,G \left( \rho_{{{\phi_0}}}+\frac12\gamma\,\rho_{{
{\gamma_0}}} \right) ^{3}}{{\Gamma }^{2} \left( \gamma-2 \right) ^{2
}{\rho_{{{\phi_0}}}}^{2}}},\label{a-end}
\end{eqnarray}
and the  e-folding number is
\begin{eqnarray}
N=\ln\left(\frac{a_{end}}{a_{begin}}\right)=\ln\left(1+\frac{16}3{\frac {\pi\,G \left( \rho_{{{\phi_0}}}+\frac12\gamma\,\rho_{{
{\gamma_0}}} \right) ^{3}}{{\Gamma }^{2} \left( \gamma-2 \right) ^{2
}{\rho^{2}_{{{\phi_0}}}}}}\right).\nonumber \\
\label{N-massless}
\end{eqnarray}

In this subsection, we got the e-folding number for interacting field between massless scalar field and radiation in super-inflation stage. According to Eq.(\ref{N-massless}),  we know   that the e-folding number  depends on the initial value of the energy density of scalar field or radiation (for $\rho_{\phi_0}+\rho_{\gamma_0}=\rho_c$) and the damping coefficient
$\Gamma$. It is easy to find that for a small damping coefficient $\Gamma$, $N$ is increasing while $\rho_{\phi_0}$ becomes bigger. Also, for a very large damping coefficient $\Gamma$, if the energy density $\rho_{\phi_0}$ is very small, it is still possible to get a enough e-folding number. But if the change of scale factor is very  large  but not  quickly enough, i.e., the duration of super-inflation is very long, then, the Taylor expanding of scale factor $a$  is not workable. To discuss the increasing the scale factor, it is necessary to use the numerical analysis.

\subsection{\label{s3B}Numerical results}

In Subsection \ref{s3A}, we discussed the e-folding number for a special case. We found that the change of the scale factor depends on the value of the damping coefficient $\Gamma$ and $\rho_{\phi_0}$, and it is possible to get enough e-folding number. But we used the Taylor series of the scale factor in this case, so the results will be very inaccurate if the scale factor changes very slow. To get a more accurate result, we will discuss the change of the scale factor with the help of the numerical method.

\begin{widetext}
\begin{center}
\begin{figure}[ht]
\begin{tabular}{cc}
\includegraphics[width=0.4\textwidth]{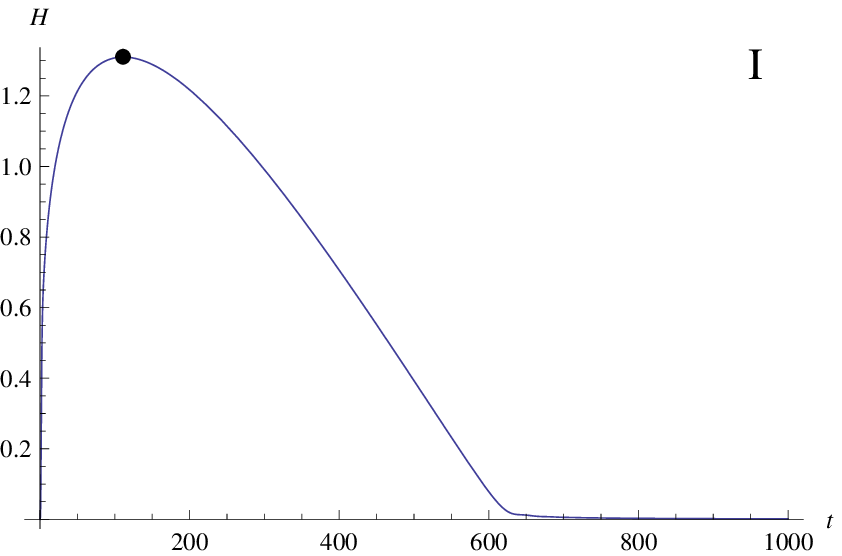}&
\includegraphics[width=0.4\textwidth]{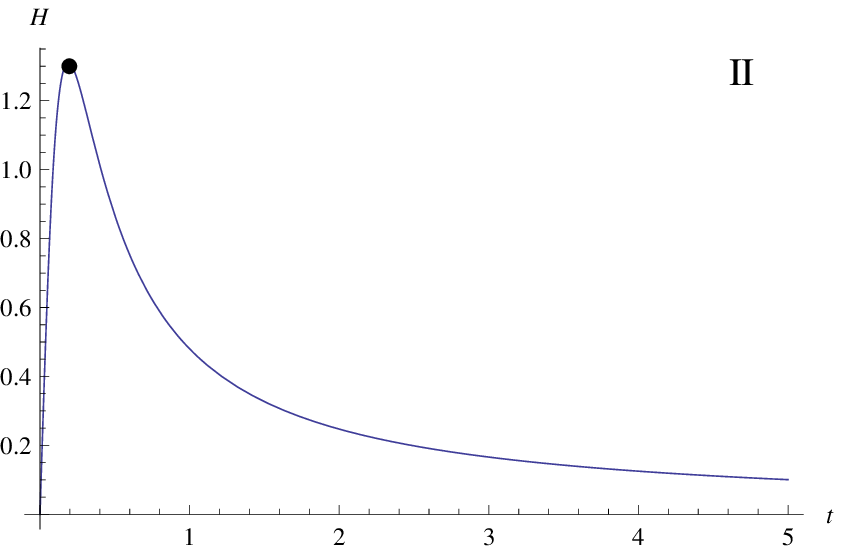}\\
\includegraphics[width=0.4\textwidth]{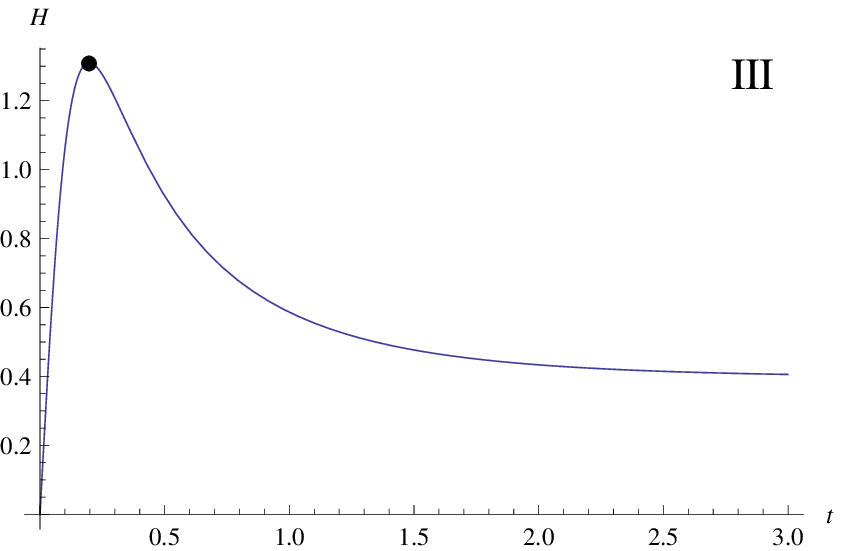}&
\includegraphics[width=0.4\textwidth]{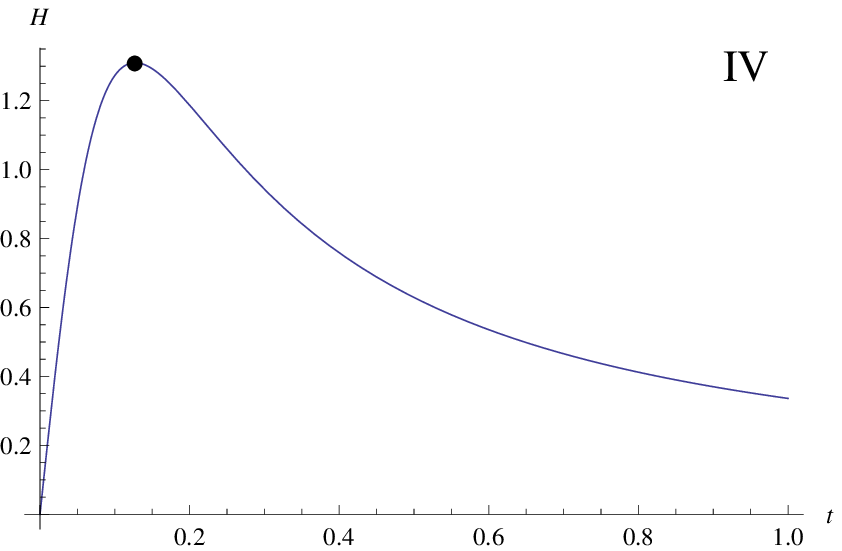}
\end{tabular}
\caption{The evolution pictures of $H$ in different cases with $\Gamma=0.01, m=0.01$. The black points show the ending time of the super-inflation. We chosen $t=0$ at the bounce point. Then we have $\dot{a}=0$ when $t=0$.  And we imaged $a=1$ at the bounce point.  The initial condition at the bounce points for scalar field and the radiation are: Case I: the radiation dominated and potential of scalar field sub-dominated ($\dot{\phi}(t=0)=0,\phi(t=0)=2,\rho_\gamma(t=0)=0.8$); Case II: the radiation dominated and the kinetic term of scalar field sub-dominated ($\dot{\phi}(t=0)=0.2,\phi(t=0)=0,\rho_\gamma(t=0)=0.8$); Case III: the kinetic term of scalar field dominated ($\dot{\phi}(t=0)\simeq 1.28,\phi(t=0)=0,\rho_\gamma(t=0)=0$) ; and  Case IV: the potential term of scalar field dominated ( $\dot{\phi}(t=0)=0, \rho_\gamma(t=0)=0, \phi(t=0)=\sqrt{2\times\frac{0.82}{0.01}}).$}
\label{fig1}
\end{figure}
\begin{figure}[ht]
\begin{tabular}{cc}
\includegraphics[width=0.4\textwidth]{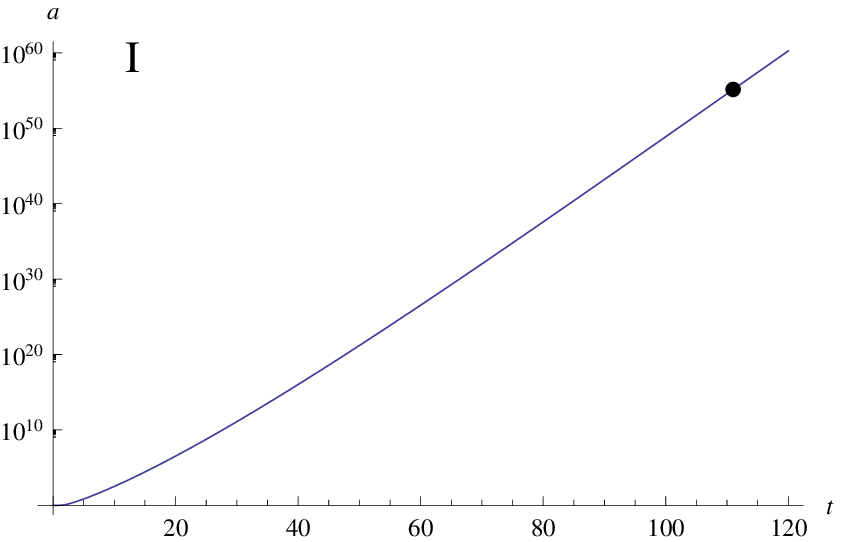}&
\includegraphics[width=0.4\textwidth]{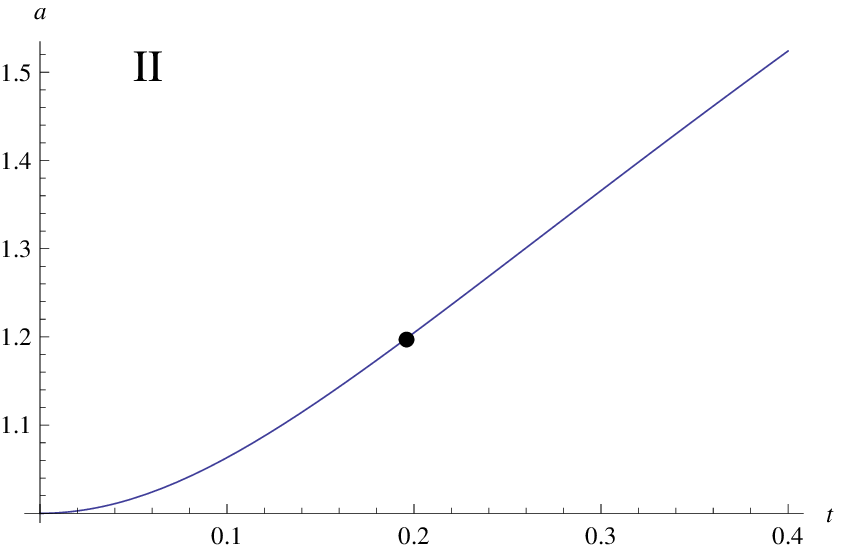}\\
\includegraphics[width=0.4\textwidth]{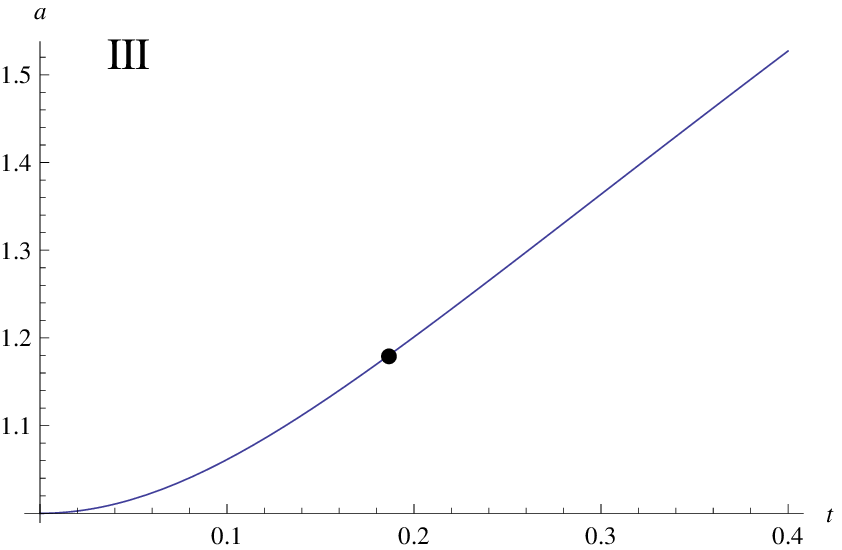}&
\includegraphics[width=0.4\textwidth]{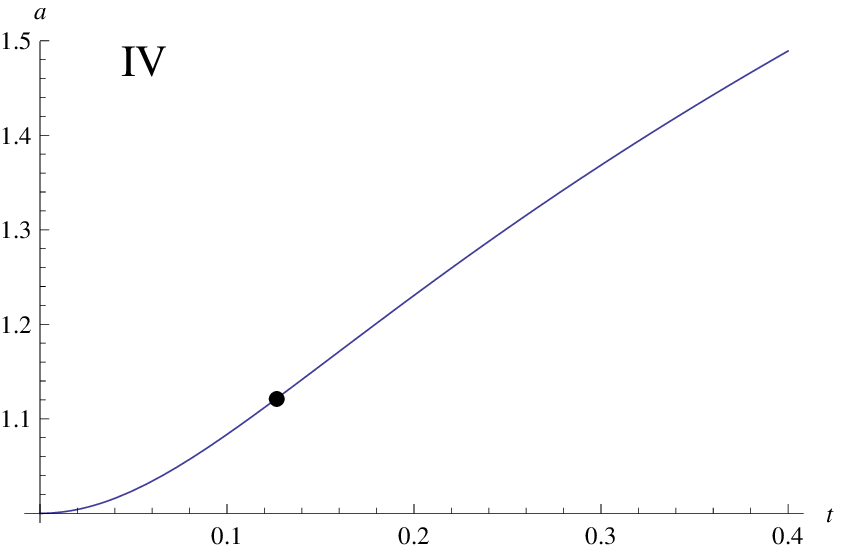}
\end{tabular}
\caption{The evolution pictures of $a$ in different cases with $\Gamma=0.01, m=0.01$. The black points show the value of the scale factor while the super-inflation ends. We chosen $t=0$ at the bounce point. Then we have $\dot{a}=0$ when $t=0$.  And we imaged $a=1$ at the bounce point.  The initial condition at the bounce points for scalar field and the radiation are as same as the values of Fig. 1.}
\label{fig2}
\end{figure}
\end{center}
\end{widetext}

To get more general results, we consider the scalar field with a quadratic potential, and the damping coefficient $\Gamma$ is a constant. At first, considering the damping coefficient $\Gamma=0.01$, we shall discuss the change of the Hubble parameter $H$ and the scale factor $a$ during the super-inflation, just as Figs.\ref{fig1} and \ref{fig2} shows. In Fig.\ref{fig1}, we showed the ending time of the super-inflation for four different examplary cases: Case I:  radiation dominates and the potential energy of scalar field sub-dominates at the bounce point; Case II, the radiation dominates and the kinetic energy of scalar field sub-dominates at the bounce point; Cases III, the kinetic energy of scalar field dominates at the bounce point; and Case IV, the potential energy of scalar field dominates at the bounce point. In Fig.\ref{fig2}, we showed the evolution pictures of the scale factor in the four cases. It is easy to see that the scale factor changes very small in the Case I, II and III, but it can increase very large during super-inflation stage if one considers the potential energy dominated at the bounce point, just as the forth picture of Fig.\ref{fig2} shows.

In Fig.\ref{fig3}, the evolution pictures of the energy density of scalar field $\rho_\phi$ and that of radiation $\rho_\gamma$ are shown. It is easy to find that the energy density of both scalar field and radiation change very small in Case I and II, whereas, in Case III and IV, the energy density of scalar field changes very slow, but the energy density of radiation changes very quickly. This is reasonable. Since in the first and second cases, radiation dominates at the bounce point, no  matter the kinetic energy or potential energy sub-dominates at the bounce point, only very small amount of energy transferred from scalar field to radiation because the duration of super-inflation is very short, just as shown in the first and second pictures of Fig.\ref{fig2}. For the third case of Fig.\ref{fig3}, the kinetic energy of scalar field dominates at the bounce point, and the energy density of radiation is increasing during super-inflation, but as same as the Case I and II, the duration of super-inflation is too short to give very obvious phenomenon of the interacting effect between scalar field and radiation. For the forth case of Fig.\ref{fig3}, the potential dominated at the bounce point. The kinetic energy of scalar field is increasing during super-inflation stage, and, at the same time, the kinetic energy of scalar field decays into radiation. Since the duration of super-inflation for the Case IV is very long, there are sufficient time for the energy of scalar field to transfer into the one of radiation. But, no matter which energy density dominates at the bounce point, the radiation will dominate at the late time.

 According to Eq.(\ref{a-end}), the evolution value of scale factor $a$ may depends on the value of the damping coefficient $\Gamma$ and the initial value of scalar field $\rho_{\phi_0}$ or radiation $\rho_{\gamma_0}$. We showed the relationship between $a$ and $\rho_{\phi_0}$ or $\rho_{\gamma_0}$ in Fig.\ref{fig1}-Fig.\ref{fig3}, now we turn to discuss the relationship between $\Gamma$ and the changes of the scale factor $a$. According to Eqs. (\ref{dphi}) and (\ref{drg}), the interaction between scalar field and radiation is the energy density of scalar field converted into that of radiation. This truth is shown by Fig.\ref{fig3}. It is easy to find that the energy density of scalar field and radiation change very small during the super-inflation stage in the Cases I, II, and III in Fig.\ref{fig3}. This means that the damping coefficient $\Gamma$ plays a very small role during the super-inflation, since there are too short time to show obvious phenomenon of the interacting effect between scalar field and radiation, no matter how big the damping coefficient $\Gamma$ is. But the energy density of radiation increasing very quickly in Case IV of Fig. \ref{fig3}, while the energy density of scalar field decreases very  slowly. This result comes from the potential domination at the bounce point, and the interacting term relates to the kinetic energy of scalar field. According to Fig.\ref{fig3}, the duration of super- inflation of Case IV is very longer than Case I, II and III, which means that there is enough time for energy density of scalar field transformed into that of radiation. Therefore, to show the relationship between $\Gamma$ and the scale factor $a$, it is nature to  choose the Case IV. We  show this in Fig. \ref{fig4}. It is easy to find that the e-folding number is increasing if one chooses a large damping coefficient $\Gamma$.
\begin{widetext}
\begin{center}
\begin{figure}[ht]
\begin{tabular}{cc}
\includegraphics[width=0.4\textwidth]{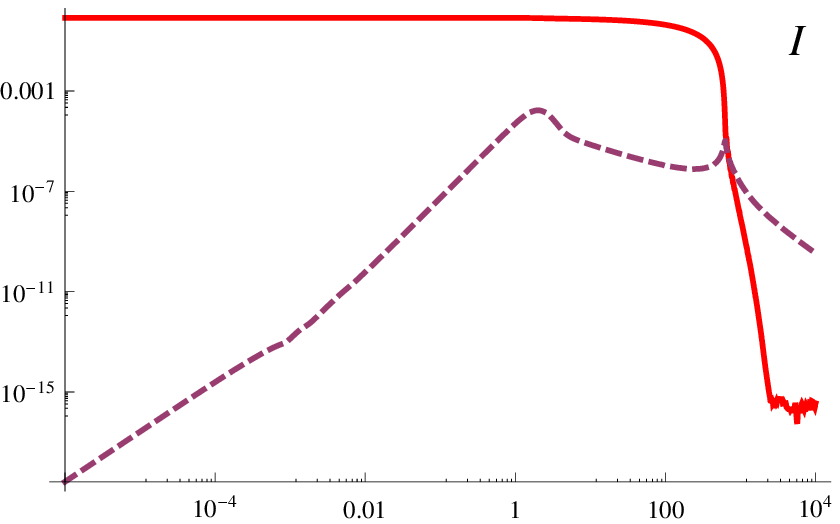}&
\includegraphics[width=0.4\textwidth]{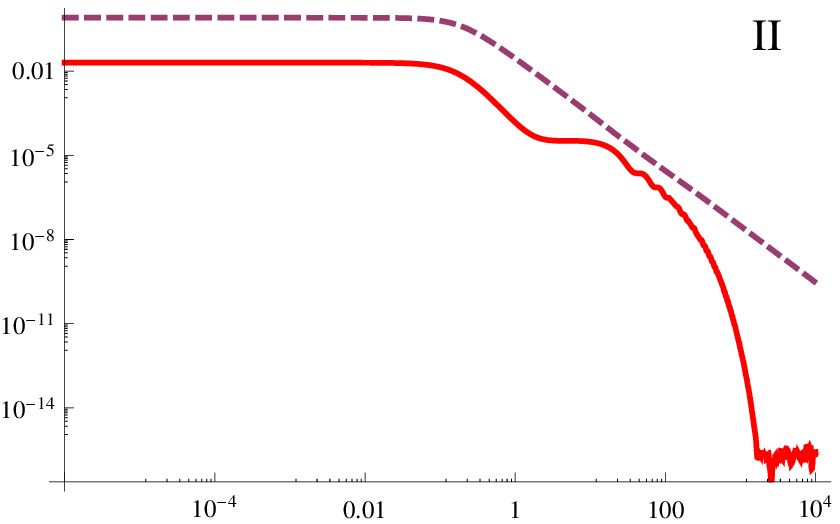}\\
\includegraphics[width=0.4\textwidth]{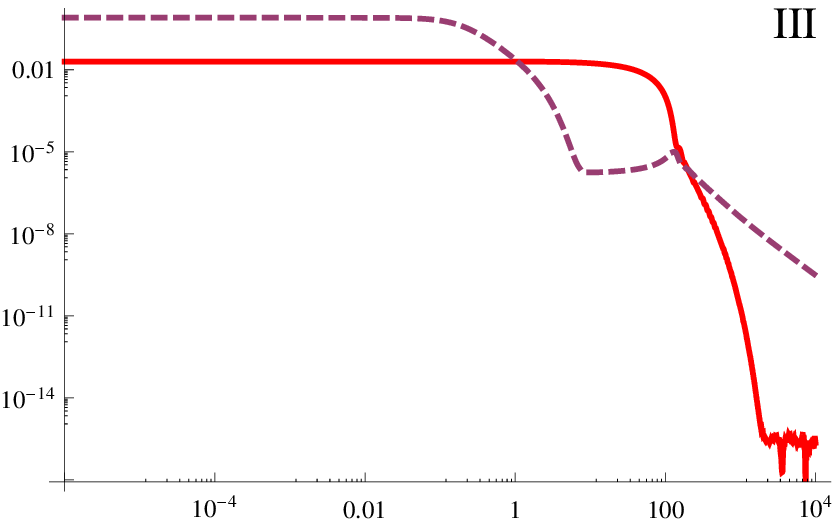}&
\includegraphics[width=0.4\textwidth]{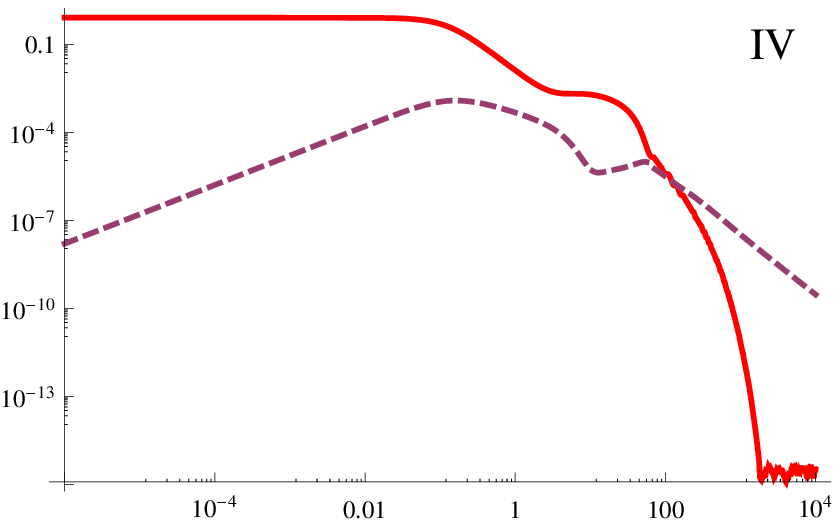}
\end{tabular}
\caption{The evolution pictures of $\rho_\phi$ (the solid line) and $\rho_\gamma$ (the dashed line) in different cases with $\Gamma=0.01, m=0.01$. We chosen $t=0$ at the bounce point. Then we have $\dot{a}=0$ when $t=0$.  And we imaged $a=1$ at the bounce point.  The initial condition at the bounce points for scalar field and the radiation are as same as the values of Fig. 1.}
\label{fig3}
\end{figure}
\begin{figure}[ht]
\begin{tabular}{cc}
\includegraphics[width=0.4\textwidth]{a-1.eps}&
\includegraphics[width=0.4\textwidth]{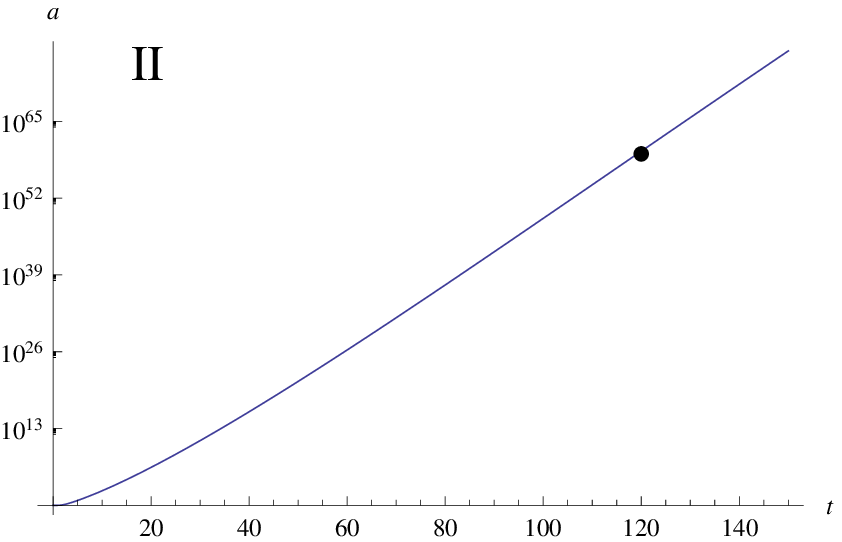}\\
\includegraphics[width=0.4\textwidth]{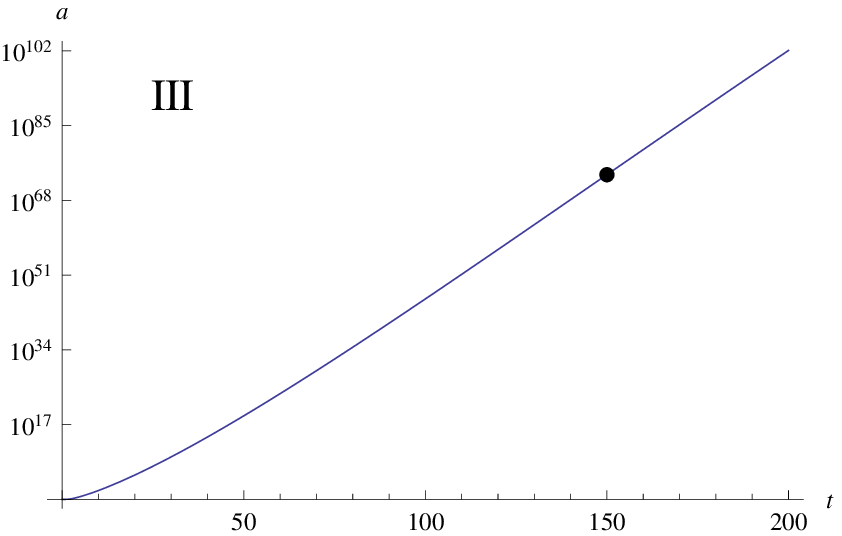}&
\includegraphics[width=0.4\textwidth]{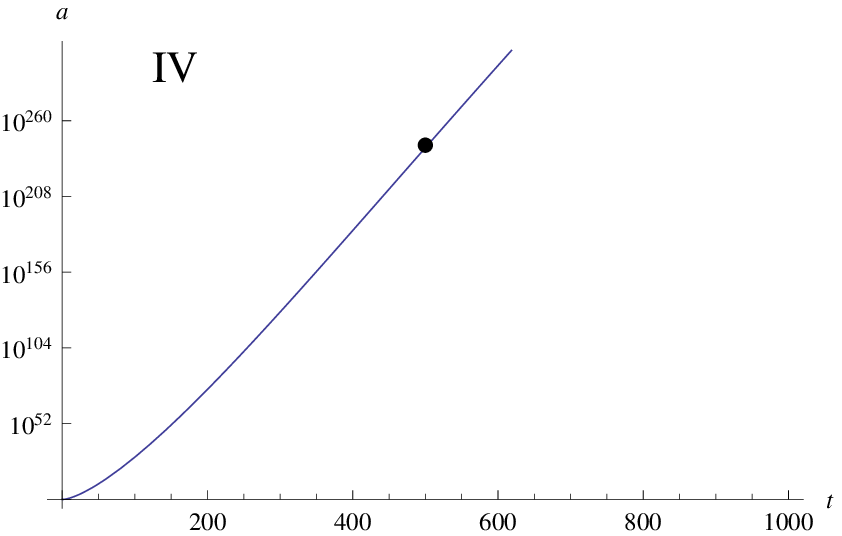}
\end{tabular}
\caption{The evolution pictures of $a$ with different $\Gamma$.  The black points show the value of the scale factor while the super- inflation ends.  Also, we choose $m=0.01$ and the bounce happens at $t=0$, and $a=1$ in the bounce point. The initial condition for scalar field and radiation as same as the Case IV we chosen in Fig. 1. The value of $\Gamma$ are $0.01, 0.1, 1, 10$ respective for I, II, III and IV. }
\label{fig4}
\end{figure}
\end{center}
\end{widetext}

\section{\label{s4}horizon problem}
In Section \ref{s3}, we have discussed the changes of scale factor during super-inflation. We found that there is just a small e-folding number during super-inflation stage except for the case that the potential dominated at the bounce point, just as we showed in Figs.\ref{fig2} and \ref{fig4}. If we follow the conclusion that the e-folding number is determined by the change of the scale factor, the results we got in the last section will be very disappointing: the super-inflation will not support enough e-folding number, and the slow rolling inflation is necessary sometimes. But it is should be remembered that the Hubble parameter is not a constant during super-inflation area \cite{Copeland-superinflation}. To solve the horizon problem, we require that $aH$ grows sufficiently during an early stage of the Universe's evolution. In fact, we need $\bar{N}\equiv \ln \frac{a(t_f)H(t_f)}{a(t_i)H(t_i)}\sim 60$ at least  \cite{Amoros-prd}. In the standard inflation theory, the Hubble parameter $H$ is nearly a constant, so the e-folding number is depending on the change of the scale factor, i.e., $\bar{N}=N\equiv\ln\frac{a_f}{a_i}$. But the Hubble parameter $H$ is not a constant during super-inflation any more \cite{Copeland-superinflation}. Following the results of Figs. \ref{fig1} and \ref{fig2}, the Hubble parameter increases very quickly while the scale factor $a$ changes very small during the super-inflation stage (just as the first, second and third pictures of the Figs. \ref{fig1} and \ref{fig2} shows), or the scale factor $a$ is changing significantly and the Hubble parameter is also increasing (just as the forth picture of Figs.\ref{fig1} and \ref{fig2} shows). There is no doubt that the super-inflation can support enough e-folding number, if one considers the potential dominated at the bounce point, just as we showed in the forth picture of Fig.\ref{fig2}. But the question is, whether the super-inflation stage can always yield enough e-folding number. As we have mentioned before, to solve the horizon problem, we need $\bar{N}\sim 60$ during super-inflation stage. According to Eq.(\ref{dH}), we know that $H=0, \dot{H}>0$ at the bounce point, this means that the Universe enter super-inflation area as soon as the quantum bounce happens. So $H(t_i)$ is the one which the value at the quantum bounce point, then $H(t_i)\simeq 0$ in LQC. It is natural to get the conclusion that $\bar{N}\equiv \frac{a(t_f)H(t_f)}{a(t_i)H(t_i)}\to\infty$ for super-inflation in LQC, so the super-inflation can solve the horizon problem. This result has been shown by many papers \cite{Copeland-superinflation, Amoros-prd}. In Ref. \cite{Copeland-superinflation},it is shown that the single scalar field can solve the horizon problem during super-inflation stage in LQC. And Ref. \cite{Amoros-prd} showed that super-inflation can support enough e-folding number while the Universe contains matter and a small cosmology constant $\Lambda$.  Also, If one considers the Universe contains a scalar field or multiple scalar field with exponential potentials, the super-inflation is also possible to yield enough e-folding number \cite{Ranken}.

To show the super-inflation stage can solve the horizon problem, we will follow the method of \cite{Amoros-prd} to calculate the particle horizon in LQC. For simplify, we choose the Case I. In this case, the radiation dominates and the kinetic energy of scalar field sub-dominates at the bounce point.It is easy to find that the radiation is always dominating during the whole super-inflation stage from the first picture of Fig. 3. According to this truth,  we assume  the energy density of scalar field is a constant $\rho_{\phi_0}=\frac12m\phi(t=0)$, and $\dot{\phi}(t)=\dot{\phi}(t=0)=0$, then using Eq.(\ref{drg}) and Eq.(\ref{Fri}) we have
\begin{eqnarray}
t &=&\frac{\sqrt{3}}4\sqrt{\frac{\rho _c}{\rho _{\phi _0}(\rho _c-\rho
_{\phi _0})}}  \nonumber \\
&&\times \ln \left\{ \frac 1{\rho _\gamma (t)}\left[ \frac{2\rho _{\phi
_0}(\rho _c-\rho _{\phi _0})+\rho _\gamma (t)(\rho _c-2\rho _{\phi _0})}{%
\rho _c}\right. \right.   \nonumber \\
&&+\frac 2{\rho _c}\sqrt{\rho _{\phi _0}(\rho _c-\rho _{\phi _0})}  \nonumber
\\
&&\left. \left. \times \sqrt{-\rho _\gamma^2 (t)+\rho _\gamma (t)(\rho
_c-2\rho _{\phi _0})+\rho _{\phi _0}(\rho _c-\phi _{\phi _0})}\right]
\right\}, \nonumber \\
\label{t-g}
\end{eqnarray}
and
\begin{eqnarray}\label{a-g}
\rho_\gamma(t) a^4(t)=\rho_\gamma(t=0)=\rho_{\gamma_0},\nonumber
\end{eqnarray}
in which the initial value $a(t=0)=1$ is used.

Following the calculation of \cite{Amoros-prd}, and with the help of Eqs.(\ref{t-g}) and (\ref{a-g}), the particle horizon
\begin{eqnarray}
d_p=a_c\int^{t_c}_{-\infty}\frac{dt}{a(t)}\nonumber
\end{eqnarray}
can be rewritten as
\begin{eqnarray}
d_p &=&\int_0^{\rho _c-\rho _{\phi _0}}\frac{\sqrt{3}}4\frac{{\rho _c}}{\rho
_{\gamma _0}^{1/4}}  \nonumber \\
&&\times \frac{d\rho _\gamma (t)}{\rho _\gamma ^{3/4}(t){\rho _c}\left[ {%
\rho _\gamma (t)+\rho _{\phi _0}}\right] {-}\left[ {\rho _\gamma (t)+\rho
_{\phi _0}}\right] ^2}  \nonumber \\
&\sim &\frac{\sqrt{3}}{4\rho _c}\int^{\rho _c-\rho _{\phi _0}}\frac{d\rho
_\gamma (t)}{\rho _c-\left[ \rho _\gamma (t)+\rho _{\phi _0}\right] }%
=+\infty
\end{eqnarray}
with $t_c, a_c$ is the bouncing time and scale factor at the bouncing point. According to the above calculation, the same result of \cite{Amoros-prd} could be obtained: \emph{when the Universe enters the expanding phase, all the points of the Universe are in causal contact and thus the Universe has had enough time to be homogeneous and isotropic when it bounces}.

\section{\label{s5}Conclusions}

As Eq.(\ref{Fri}) shows, the Friedmann equation in LQC adds a $(1-\rho/\rho_c)$ term in the right side of the standard Friedmann equation. The correction term $\rho/\rho_c$ comes with a negative sign, this makes the possibility that $\dot{a}=0$ when $\rho=\rho_c$, and the bounce occurs. At the bounce point, $H=0$ and $\dot{H}$ is positive and the Universe enters a super-inflation stage. Equation (\ref{dH}) shows that $\dot{H}$ continues to positive till $\rho=\frac12\rho_c$, at which point it vanishes (after this point, it will become negative.). Thus, every LQC solution has a super-inflation phase from $\rho=\rho_c$ to $\rho=\frac12\rho_c$, which is staying in the quantum geometry dominated area. Lots of papers researched the characteristic of super-inflation \cite{Bojowald-PRL-in,Copeland-superinflation,Ranken,Ribassin,Amoros-prd}. some concluded that the suer inflation can not solve the horizon problem, while others argued it is possible. All these papers considered that the Universe fills with a massless or self-interacting scalar field, or matter with a small cosmology constant $\Lambda$. In this Letter, we discuss the super- inflation in the condition that the contains a interacting scalar field and radiation.

In the standard inflation theory, the e-folding number comes from the increment of the scale factor $a$ during inflation stage. To check the increasing value of the scale factor during super-inflation area, we considered the massless scalar field interacts with radiation at first. Assumed the duration of super-inflation is very short and $\Gamma\gg H$, we expanded the scale factor and got the expression of the e-folding number. We found that the e-folding number depends on the value of $\Gamma$ and $\rho_{\gamma_0}$ (or $\rho_{\phi_0}$ for $\rho_{\gamma_0}+\rho_{\phi_0}=\rho_c$ at the bounce point), and it is possible to get a very large change of the scale factor, or get a very large value of e-folding number. Since we used the Taylor expansion of the scale factor, this result is not reliable for all cases. To get more reliable result, we using the numerical method to analyze the evolutions picture of Hubble parameter $H$, the scale factor $a$, and the energy density $\rho_\phi,\rho_\gamma$. Four cases are considered, just as we mentioned in Fig. 1. We find that the scalar field changes very small in the first three cases, but changes very drastically in the forth case. For all four cases, the Hubble parameter changes during super-inflation stage for $\dot{H}>0$. Also,  we
find the value of  $\Gamma$ also affects the value of the scale factor in the Case IV, but it has very little influence on the first three cases, for the duration of super-inflation is too
short to accumulate the interacting effect  between the scalar field and radiation. Moreover, we find that the Universe will finally enter a radiation dominated stage, no matter which energy density dominated at the bounce point.

The Hubble parameter changes in the super-inflation stage, so the e-folding number depends not only on the variation of scale factor, but also on the Hubble parameter. Just as \cite{Amoros-prd} argued, the e-folding number is $\bar{N}\equiv \ln\frac{a(t_f)H(t_f)}{a(t_i)H(t_i)}\sim 60$ during super-inflation area. According to Eqs.(\ref{Fri}) and Eq.(\ref{dH}), it is easy to find that the Universe enter the super-inflation stage when the bounce happens, which means that  the initial value of Hubble parameter $H(t_i)\sim 0$, then $\bar{N}\sim \infty$ at the super-inflation stage. To show that it is possible to solve the horizon problem in super-inflation stage, we followed the method of \cite{Amoros-prd} to calculate the particle horizon. Assuming the radiation dominates at the bounce point, we find that $d_p=\infty$, then all the points of the Universe are in causal contact.

To obtain more information of super-inflation in LQC, perturbation theory should be considered. The perturbation theory of effective LQC has been discussed by many papers, and the possible observational effect has been analyzed (more recently review, please see \cite{Ivan,Florian,Gianluca}). In this Letter, we just discuss some phenomenological results of super-inflation for the case that the Universe is filled with interacting scalar field and radiation. The perturbation theory of interacting scalar field and radiation in LQC still needs to be  researched, we will back to this issue very soon.

\acknowledgments Zhu was supported by the National Natural Science Foundation of China (Grant Nos. 11175019 and 11235003) and Xiao was supported by the National Natural Science Foundation of China (Grant Nos. 11175019 and 11247282).


\begin{thebibliography}{111}
\bibitem{Lindle}David H. Lyth, and Andrew R. Liddle, \emph{The Primordial Density Perturbation: Cosmology, Inflation and the Origin of Structure}, Cambridge University Press, Cambridge, 2009.
\bibitem{Lisa-warm}Sam Bartrum, Arjun Berera, Joao G. Rosa, JCAP \textbf{1306}, 025(2013).
\bibitem{inflation-planck} Planck Collabration: R. A. R. Ade, \emph{et al}, arXiv: 1303.5082 [astro-ph.Co].
\bibitem{Ivan-LQC}Ivan Agullo, Alejandro Corichi, arXiv: 1302.3833[gr-qc].
\bibitem{Bojowald-book}
M. Bojowald, \emph{Quantum Cosmology: A fundamental description of the Universe}, Lect. Notes. Phys. \textbf{835}, 1 (2011)
\bibitem{Rovelli-Book} C. Rovelli, \emph{Quantum Gravity}, Cambridge university press, Cambridge, 2004.
\bibitem{Thiemann-Book} T. Thiemann, \emph{Modern Canonical Quantum General Relativity}, Cambridge university press, Cambridge, 2007.
\bibitem{Ashtekar} A. Ashtekar, T. Pawlowshik, and P. Singh, 
  Phys. Rev. Lett. \textbf{96}, 141301(2006). 
  Phys. Rev. D \textbf{73}, 124038(2006).
\bibitem{Singh} P. Singh and A. Toporensky, 
  Phys. Rev. D \textbf{69}, 104008(2004).
  G.V. Vereshchagin, 
  J. Cosmol. Astropart. Phys.  \textbf{0407}, 013(2004). G. Date and G.M. Hossain, 
  Phys. Rev. Lett. \textbf{94}, 011302(2005).
\bibitem{Haro-JCAP}Kauzuharu Bamba, Jaume de Haro, Sergei D. Odintsov, JCAP, 1302, 008 (2013).
\bibitem{Ashtekar-probability}A. Ashtekar, D. Sloan, Gen. Rel. Grav. \textbf{43}, 3619 (2011).
\bibitem{Corichi_measure}A. Corichi, A. Karami, Phys. Rev. D \textbf{83}, 104006 (2011).
\bibitem{Linsefors-prediction} L. Linsefors, A. Barrau, Phys. Rev. D \textbf{87}, 123509 (2013).
\bibitem{Bojowald-PRL-in} M. Bojowald, 
  Phys. Rev. Lett. \textbf{89}, 261301(2002). M. Bojowald and K. Vandersloot,  
  Phys. Rev. D 67 (2003) 124023.
\bibitem{Copeland-superinflation} E.J. Copeland, D.J. Mulryne, N.J. Nunes, and M. Shaeri, 
  Phys. Rev. D \textbf{77}, 023510(2008).
\bibitem{Ranken}E. Ranken, P. Singh, Phys. Rev. D \textbf{85}, 104002 (2012).
\bibitem{Ribassin} J. Ribassin, E. Huguet, and K. Ganga, arXiv:1111.4661[gr-qc].
\bibitem{Amoros-prd}Jaume Amoros, Jaume de Haro, and Sergei D. Odintsov, Phys. Rev. D \textbf{87}, 104037 (2013).
\bibitem{Mielczarek}J. Mielczarek, T. Cailleteau, J. Grain, A. Barrau, Phys. Rev. D \textbf{81}, 104049 (2010).
\bibitem{Herrera} R. Herrera, Phys. Rev. D \textbf{81}, 123511 (2010).
\bibitem{Xiao} K. Xiao and J.Y. Zhu, Phys. Lett. B \textbf{699}, 217 (2011).
\bibitem{Zhang} Xiao-Min Zhang, Jian-Yang Zhu, Phys. Rev. D \textbf{87}, 043522 (2013).
\bibitem{Edward} Edward Wilson-Ewing, JCAP \textbf{1303}, 026(2013).
\bibitem{Ivan} Ivan Agullo, arXiv: 1302.3833[gr-qc].
\bibitem{Florian}F. Girelli, F. Hinterleitner and Seth A. Major, AIGMA \textbf{8}, 098 (2012).
\bibitem{Gianluca}G. Calcagni, Ann. Phys. \textbf{525}, 323 (2013).

\end{thebibliography}
\end{document}